\documentclass[final,3p,times]{elsarticle}

\usepackage{amssymb}
\usepackage{amsmath}

\begin{document}

\begin{frontmatter}



\title{Maximal mass of neutron stars constrained by neutron star observations}


\author{G\'abor Kasza}
\ead{kasza.gabor@wigner.hun-ren.hu}
\affiliation{organization={Theory Department, HUN-REN Wigner RCP},
            addressline={29-33 Konkoly-Thege}, 
            city={Budapest},
            postcode={1121}, 
            country={Hungary}}
\affiliation{organization={Institute of Technology, MATE},
            addressline={36 Mátrai út}, 
            city={Gyöngyös},
            postcode={3200}, 
            country={Hungary}}

\author{Gy\"orgy Wolf}
\ead{wolf.gyorgy@wigner.hun-ren.hu}

\affiliation{organization={Theory Department, HUN-REN Wigner RCP},
            addressline={29-33 Konkoly-Thege}, 
            city={Budapest},
            postcode={1121}, 
            country={Hungary}}

\begin{abstract}
We investigate constraints on the high-density equation of state (EOS) of neutron star matter by analyzing the probability distributions of the endpoints of mass–radius M(R) sequences within a Bayesian weighting framework. Starting from two representative hadronic baseline EOSs, SFHo and DD2, matched at higher densities to an extended linear sigma model description and constrained to approach perturbative QCD (pQCD) results, we construct families of causal hybrid EOSs spanning a broad range of stiffness at supranuclear densities. Observational constraints from the binary neutron-star merger GW170817, mass–radius measurements from the Neutron Star Interior Composition Explorer (NICER), and candidate low-mass and mass-gap compact objects are incorporated through Bayesian likelihood weighting. This approach allows us to determine probability distributions for the maximum neutron-star mass M$_{\rm TOV}$ and the corresponding radius R$_{\rm TOV}$, i.e., the endpoints of the M(R) sequences.

We find that the maximum-mass distributions are largely determined by observational constraints and show only weak sensitivity to the choice of baseline EOS, favoring values around 2.2–2.3 M$_\odot$ when the most robust constraints are applied. In contrast, the corresponding radius distributions exhibit a stronger dependence on the underlying hadronic EOS, with typical preferred values near
$12\pm 1$ km. Additional tidal-deformability constraints further restrict the allowed parameter space and disfavor very stiff EOS realizations when interpreted together with the possible mass-gap neutron-star candidate. Our results demonstrate that endpoint distributions of M(R) sequences provide a sensitive and complementary diagnostic for constraining the high-density behavior of the neutron-star EOS within a multimessenger Bayesian framework.
\end{abstract}




\end{frontmatter}



\section{Introduction}
\label{sec1}

One of the central goals of heavy-ion and nuclear physics is to determine the properties and phase diagram of strongly interacting matter. Experimentally, this matter can presently be explored only in limited regions of the phase diagram: at very small baryon chemical potential—accessible at facilities such as Relativistic Heavy Ion Collider and Large Hadron Collider—and around nuclear saturation density in conventional nuclear-structure experiments. In the low-chemical-potential regime, lattice QCD provides reliable first-principles results, while at extremely large chemical potentials perturbative QCD (pQCD) becomes applicable. However, at intermediate chemical potentials—where the critical end point is expected to reside—the properties of strongly interacting matter remain largely unknown both experimentally and theoretically.

Future accelerator facilities such as NICA and FAIR are expected to probe dense strongly interacting matter at temperatures of order $T\sim70$–$120$ MeV. In contrast, no terrestrial experiments can access strongly interacting matter at high density and nearly zero temperature. At present, neutron stars provide the only known environment where matter reaches several times nuclear saturation density under such conditions. Observations of neutron stars therefore offer a unique opportunity to constrain the equation of state (EOS) of cold dense matter through electromagnetic, gravitational-wave, and multi-messenger measurements.

At zero density, cold strongly interacting matter is constrained by nucleon–nucleon scattering data. Around nuclear saturation density $\rho_0$, a wide range of nuclear-structure observables—including nuclear masses, isobaric analog states, hypernuclei, giant dipole and pygmy resonances, dipole polarizability, and neutron-skin thickness—provides strong constraints on the EOS up to approximately $\rho_0$ and somewhat beyond. At densities $\rho\lesssim1.5\rho_0$, hadronic effective field theories further provide reliable description of nuclear matter and nucleon–nucleon interactions.

At the opposite end of the density range, corresponding to $\rho\sim40\rho_0$, pQCD calculations determine the pressure at $\mu\simeq2.6$ GeV to be approximately $3.8\:\mathrm{GeV/fm^3}$ of $\mathcal{O}(\alpha_s^3\ln\alpha_s)$ order within the hard-thermal-loop framework. Although such densities are not realized in neutron stars, these results provide an important asymptotic constraint: any realistic EOS must connect smoothly to the pQCD regime while remaining causal, i.e., with the speed of sound below the speed of light.

In this work, we investigate how neutron-star observations constrain the EOS of strongly interacting matter by constructing a unified description of hadronic and quark matter  (see details in \citep{Kasza-2026}). For the low-density regime we employ the relativistic mean-field model of Steiner et al., (SFHo model \citep{Hempel2009,Steiner2013} ), which includes nucleons interacting via $\sigma$, $\omega$, and $\rho$ mesons with nonlinear couplings. This EOS is relatively soft, with incompressibility $K=245$ MeV, slope of the symmetry-energy $L=47.1$ MeV, and effective-mass ratio $m^\star/m_n=0.76$. 
As an alternative for hadronic matter we use the DD2 hadronic EOS~\cite{Hempel:2009mc,Typel:2009sy}, which is stiffer than the SFHo EOS.

At higher densities we employ the quark–meson model based on $U(3)\times U(3)$ chiral symmetry, which includes constituent quarks, (pseudo)scalar and (axial-)vector meson nonets, and Polyakov loops. The model parameters are fixed by vacuum hadron properties and reproduce lattice-QCD thermodynamics at vanishing chemical potential, remaining consistent with lattice results up to $\mu\lesssim400$ MeV. In the present work we adopt the parameter set with $m_\sigma=290$ MeV.

To construct a unified EOS across the full density range relevant for neutron stars, the low- and high-density regimes must be connected. Because the two models involve different degrees of freedom and have limited domains of validity, an interpolation is required at intermediate densities. Astrophysical observations strongly constrain the presence of large first-order phase transitions in this region, motivating a smooth matching procedure. Following earlier work, we interpolate between the two regimes using a polynomial ansatz for the energy density as a function of baryon density. This choice ensures thermodynamic consistency and guarantees continuity of both the pressure and the speed of sound across the transition region. Specifically, assuming that the hadronic EOS $\varepsilon_{\mathrm H}(n_B)$ is valid up to $\rho_{BL}$ and the quark EOS $\varepsilon_{\mathrm Q}(n_B)$ above $\rho_{BU}$, we construct a fifth-order polynomial interpolation in the interval $\rho_{BL}<\rho_B<\rho_{BU}$ constrained to ensure continuity of the energy density and its first two derivatives at the matching points.

Finally, we assume that the quark–meson EOS remains applicable up to the highest densities realized in neutron-star cores and can be smoothly connected to perturbative QCD results in the asymptotic high-density limit. This framework allows us to explore how present neutron-star observations constrain the EOS of cold strongly interacting matter across a wide density range.

\section{Observational constraints}
\label{sec:data}

Since in this paper we investigate the probability distribution of the maximal mass of neutron stars, therefore, we do not use here the maximum mass constraints, like in ref. \citep{Kasza-2026}.
\begin{itemize}

\item \textbf{Perturbative-QCD constraint.}

At asymptotically high densities the EOS must approach perturbative-QCD results while remaining causal ($c_s<1$) and thermodynamically consistent, ensuring stability and allowing a causal interpolation to the pQCD regime. Following the N$^3$LO calculation of Gorda \textit{et al.} \cite{Gorda2018,Gorda2021,Gorda2023}, we require consistency with the reference point
\[
\mu_{\mathrm{QCD}}=2.6~\mathrm{GeV}, \quad
\rho_{\mathrm{QCD}}=6.47~\mathrm{fm^{-3}}, \quad
p_{\mathrm{QCD}}=3823~\mathrm{MeV/fm^3}.
\]

\item \textbf{NICER mass–radius measurements.}

Pulse-profile modeling of surface hot spots observed by the Neutron Star Interior Composition Explorer (NICER) provides simultaneous constraints on neutron-star masses and radii with $\sim10\%$ precision. In this work we include recent NICER results for five pulsars:
PSR~J0030+0451,
PSR~J0740+6620,
PSR~J0614–3329,
PSR~J1231–1411,
and PSR~J0437–4715,
using the latest available posterior distributions from Refs.~\citep{Vinciguerra2024,Salmi:2024aum,Mauviard2025,Salmi2024,Choudhury2024}.

\item \textbf{GW170817 tidal-deformability constraint.}

The first binary neutron-star merger detection, GW170817, provided important constraints on the EOS through tidal deformability. We adopt the conservative EOS-agnostic bound
\[
\tilde{\Lambda}<720
\]
from the LIGO–Virgo analysis \citep{LIGOScientific:2018hze}. We also examine the impact of the stronger constraint (``$\Lambda_{190}$'') 
\[
70<\Lambda_{1.4}<580,\qquad
9.1~\mathrm{km}<R_{1.4}<12.8~\mathrm{km},
\]
derived from combined tidal-deformability and radius measurements \citep{LIGOScientific:2018cki}. The intervals shown correspond to 90\% confidence intervals.

\item \textbf{Low-mass compact-object candidate (HESS J1731–347).}

Observations of the central compact object in HESS~J1731–347 suggest a possible very low-mass neutron star with
\[
M=0.77^{+0.20}_{-0.17}\,M_\odot,\quad
R=10.4^{+0.86}_{-0.78}\,\mathrm{km}.
\]
Because the interpretation of this object remains uncertain \citep{Sagun:2023rzp}, we treat this constraint with caution.

\item \textbf{Mass-gap object from GW190814.}

The compact object detected in GW190814 has inferred mass
\[
M=2.59^{+0.08}_{-0.09}\,M_\odot,
\]
placing it in the lower mass gap between neutron stars and black holes \citep{LIGOScientific:2020zkf,Mali2026}. We explore the implications of interpreting this object as a neutron star by applying a Gaussian constraint centered at $2.59\,M_\odot$ with width $\Delta M=0.055\,M_\odot$.
\end{itemize}

\section{Results}

\begin{figure}[t]
\centering
\includegraphics[width=0.45\textwidth]{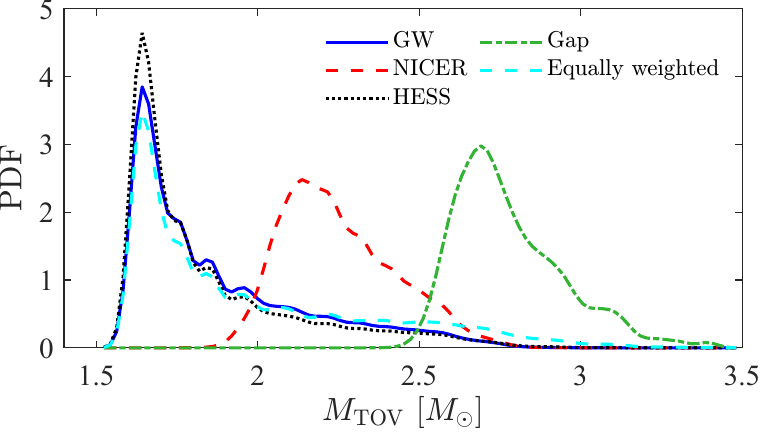}
\includegraphics[width=0.45\textwidth]{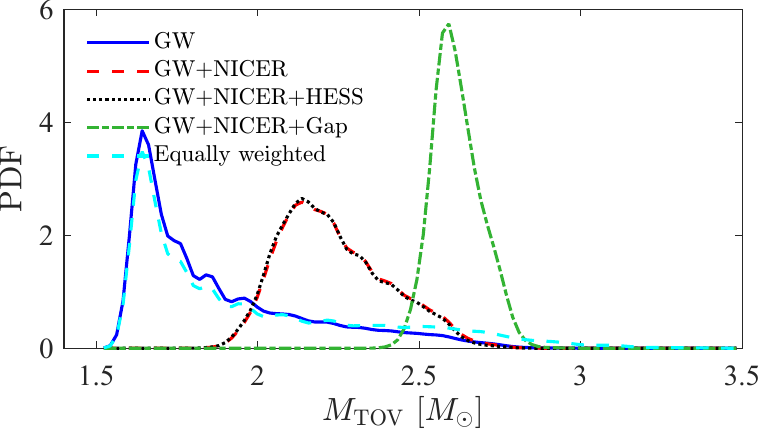}\\[0.2cm]

\includegraphics[width=0.45\textwidth]{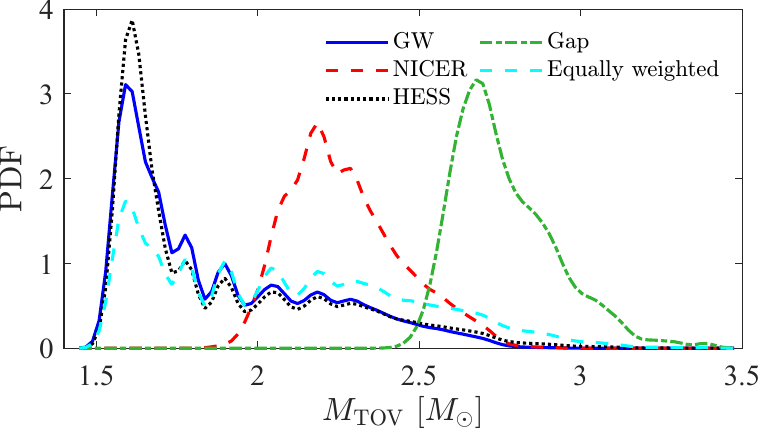}
\includegraphics[width=0.45\textwidth]{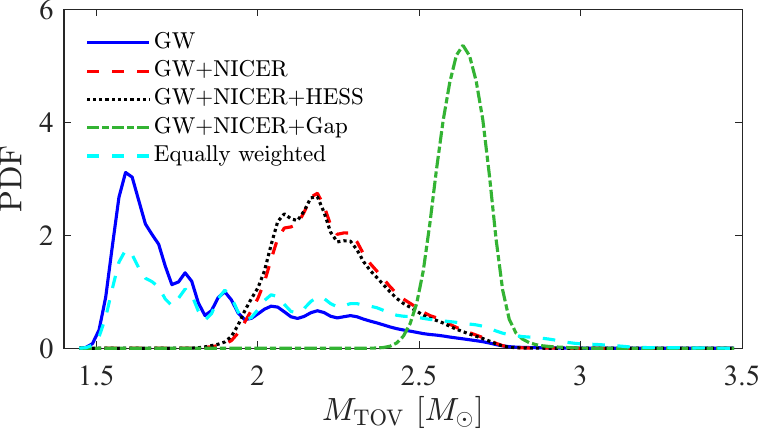}

\caption{Probability distribution of maximal mass of neutron stars using different neutron star observations. The pQCD constraint were applied in all the figures. The upper panels were calculated with the SFHo+eLSM EOS, while the ones below the SFHo were exchanged to DD2.}\label{fig:Maxmass}
\end{figure}
 
\begin{figure}[t]
\centering
\includegraphics[width=0.45\textwidth]{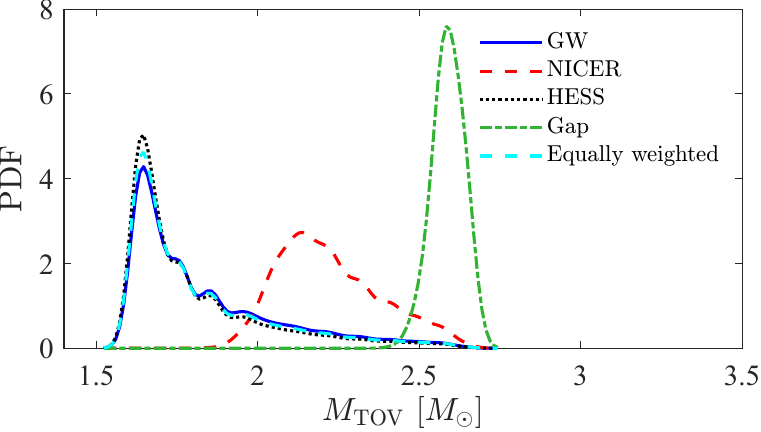}
\includegraphics[width=0.45\textwidth]{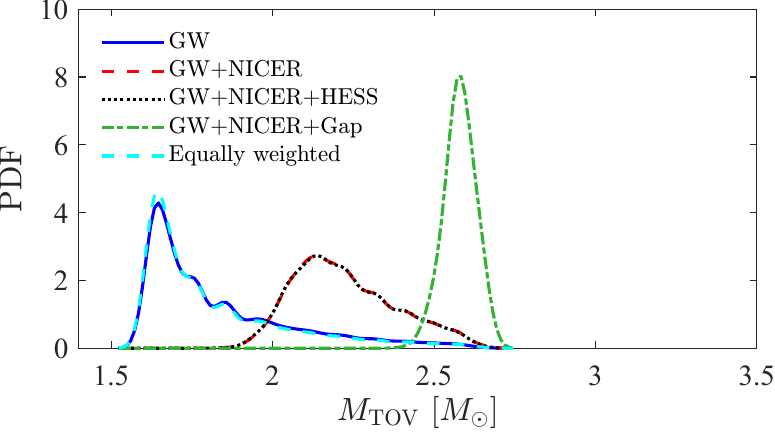}\\[0.2cm]

\includegraphics[width=0.45\textwidth]{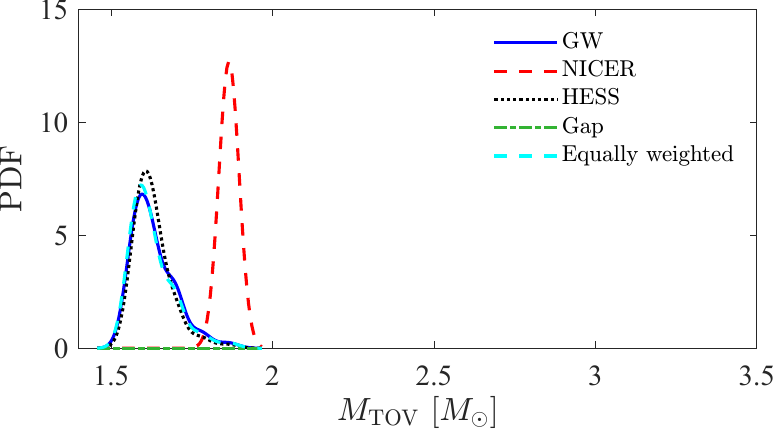}
\includegraphics[width=0.45\textwidth]{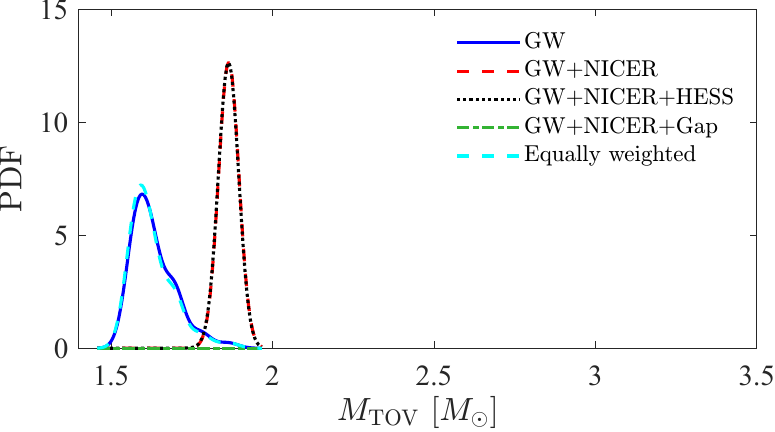}
\caption{The same as Fig.~\ref{fig:Maxmass}, but adding the constraint of $\Lambda_{190}$.}\label{fig:Maxmasslam190}
\end{figure}

\begin{figure}[t]
\centering
\includegraphics[width=0.45\textwidth]{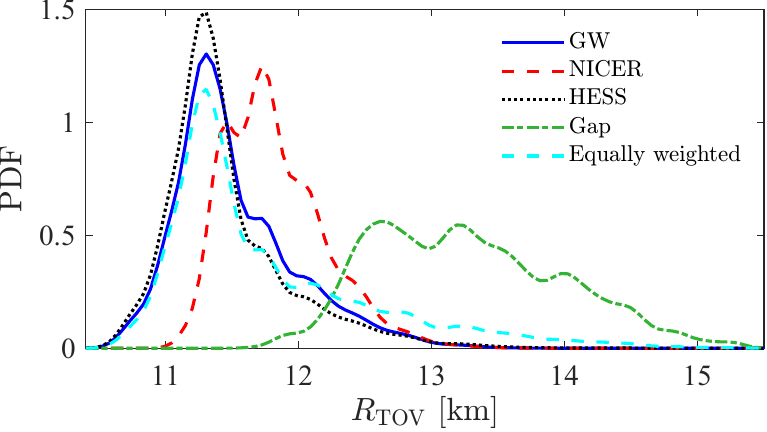}
\includegraphics[width=0.45\textwidth]{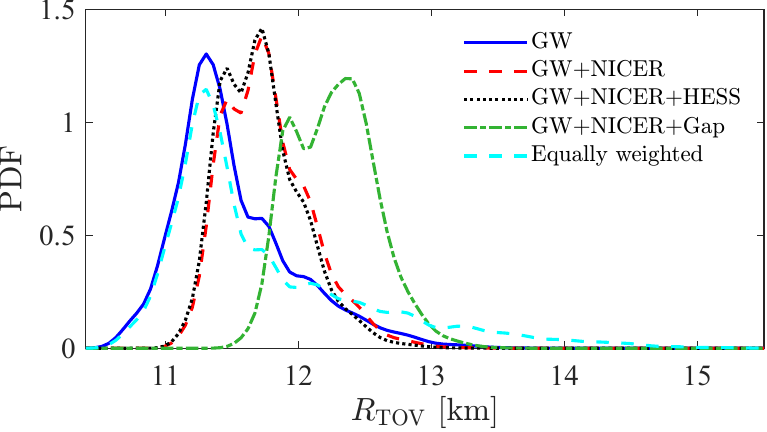}\\[0.2cm]

\includegraphics[width=0.45\textwidth]{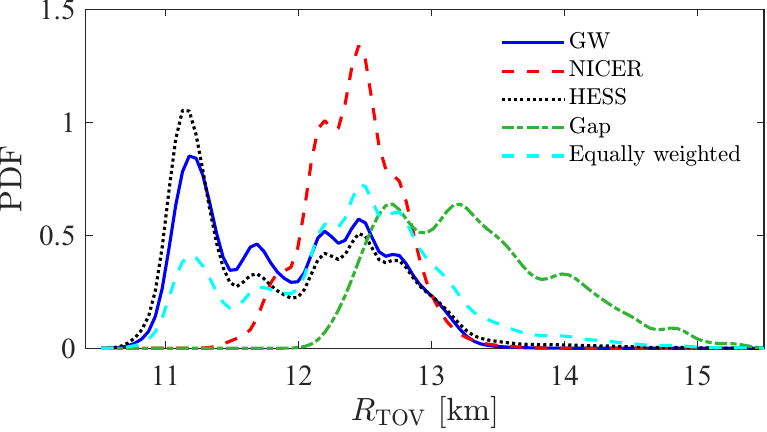}
\includegraphics[width=0.45\textwidth]{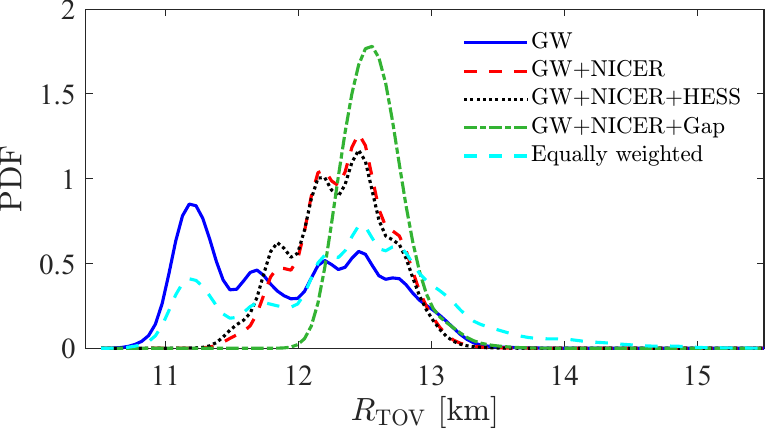}
\caption{Probability distribution of radii corresponding to the maximal mass of neutron stars using different neutron star observations. The pQCD constraint were applied in all the figures. The upper panels were calculated with the SFHo+eLSM EOS, while the ones below the SFHo were exchanged to DD2.}\label{fig:RTOV}
\end{figure}

\begin{figure}[t]
\centering
\includegraphics[width=0.45\textwidth]{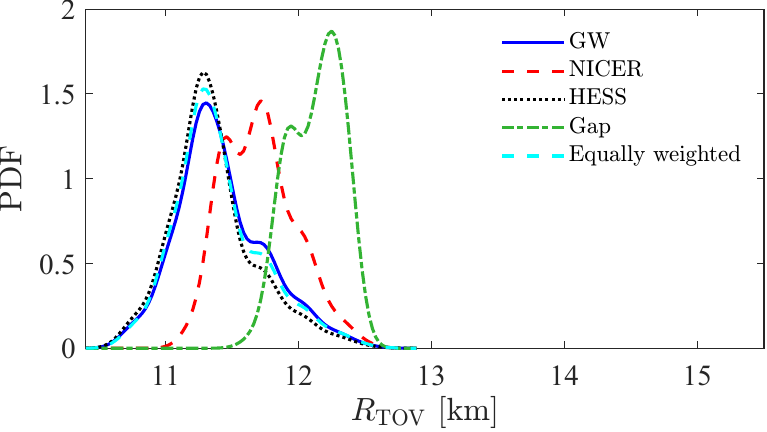}
\includegraphics[width=0.45\textwidth]{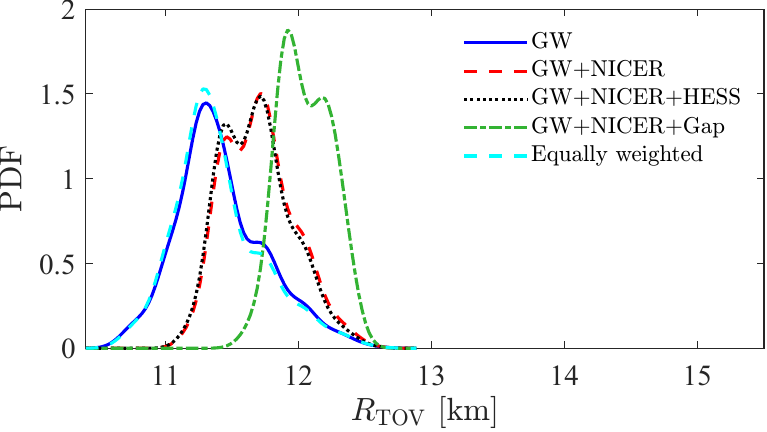}\\[0.2cm]

\includegraphics[width=0.45\textwidth]{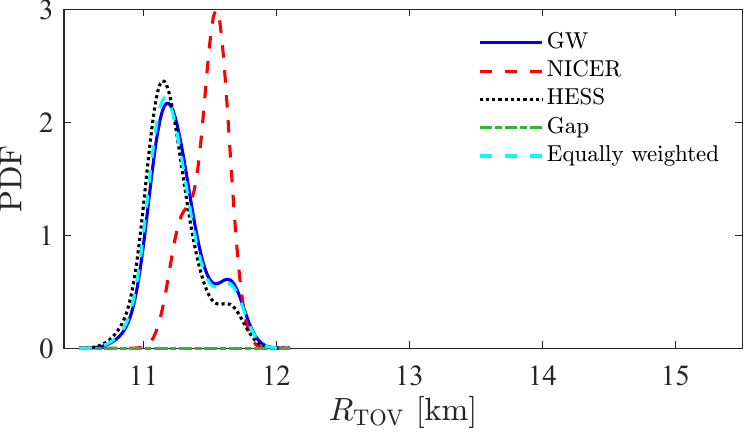}
\includegraphics[width=0.45\textwidth]{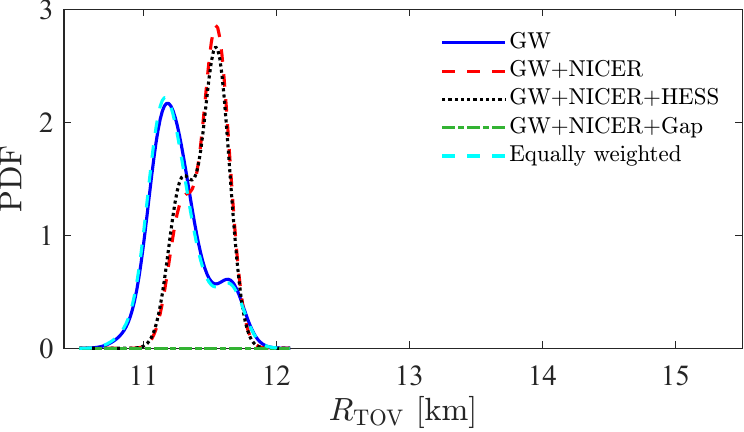}
\caption{The same as Fig.~\ref{fig:RTOV}, but adding the constraint of $\Lambda_{190}$.}\label{fig:RTOVlam190}
\end{figure}
 
We use Bayesian inference to investigate the impact of the observational data described in Sec. \ref{sec:data} on the manifold of possible equations of state (EOSs) (see details in \citep{Kasza-2026}).

We investigated the probability distributions of the endpoints of the M(R) sequences, namely the maximum neutron star mass M$_{\rm TOV}$ and the corresponding radius R$_{\rm TOV}$. The analysis was performed by first applying the pQCD constraint and then sequentially adding observational constraints. The resulting distributions obtained with the SFHo+eLSM and DD2+eLSM equations of state (EOSs) are shown in Fig.~\ref{fig:Maxmass}.

Using only the pQCD constraint already produces a relatively well-defined maximum-mass distribution, with a peak slightly above 1.5 M$_\odot$. Including the GW170817 or HESS constraints does not significantly modify the peak position, although moderate changes in the width of the distributions are visible, particularly for the DD2+eLSM EOS. The inclusion of the NICER constraint shifts the peak toward 2.1–2.2 M$_\odot$ and leads to a visibly sharper distribution. As expected, imposing the mass-gap constraint results in a pronounced shift of the peak toward larger values, around 2.6–2.7 M$_\odot$. Overall, both EOS constructions lead to qualitatively similar probability distributions for the maximum mass, indicating that the location of the peak is primarily driven by observational constraints rather than by the choice of the low-density baseline EOS.

If the additional tidal-deformability constraint ``$\Lambda _{190}$'' is imposed (Fig.~\ref{fig:Maxmasslam190}), the results obtained with the SFHo-based EOS remain qualitatively unchanged compared to the case without this constraint. In contrast, for the DD2-based EOS the inclusion of this constraint prevents a consistent description of the mass-gap neutron-star candidate. This reflects the comparatively large stiffness of the DD2 EOS, which leads to tidal deformabilities that are systematically too large to satisfy the imposed ``$\Lambda_{190}$'' interval while simultaneously supporting such high maximum masses.

We also analyzed the probability distributions of the radii corresponding to the endpoints of the M(R) sequences (Fig.~\ref{fig:RTOV}). In contrast to the maximum-mass case, here the SFHo+eLSM and DD2+eLSM EOSs exhibit noticeable qualitative differences.

For the SFHo+eLSM EOS, the pQCD constraint alone—and its combination with either the GW170817 or HESS constraints—produces a stable distribution peaked near 11.2 km. Adding the NICER constraint shifts the peak to approximately 11.8 km, while including the mass-gap constraint results in a broader distribution centered around 13 km.

For the DD2+eLSM EOS, the situation is different. The pQCD, pQCD+GW170817, and pQCD+HESS constraints yield a primary peak slightly above 11 km, together with a secondary structure in the range 12–13 km. The NICER constraint strengthens the larger-radius contribution and produces a pronounced peak near 12.4 km. When the mass-gap constraint is included, the resulting distribution becomes broader and resembles the behavior observed in the SFHo+eLSM case.

If the $\Lambda_{190}$ constraint is also applied, the distributions of $R_{\rm TOV}$ become narrower, independently of which observational constraint is used to weight the samples (Fig.~\ref{fig:RTOVlam190}). This result is consistent with expectations, given the relation between $\Lambda$ and the radius: constraining the tidal deformability to a lower range leads to a decrease in the expected radius.

When all well-established constraints (pQCD + GW170817 + NICER) are applied simultaneously, both EOSs predict a maximum-mass distribution peaked at 2.2–2.3 M$_\odot$. The corresponding radius distributions peak at approximately 11.8 km for SFHo+eLSM and 12.4 km for DD2+eLSM, indicating a residual but systematic dependence on the underlying hadronic baseline EOS.

\section{Summary}

In summary, the probability distributions of the maximum neutron star mass M$_{\rm TOV}$ are largely controlled by observational constraints and show only a weak dependence on the choice of baseline EOS. The pQCD constraint already favors values slightly above 2.1 M$_\odot$, while the inclusion of GW170817 and HESS constraints produces only moderate modifications. NICER measurements sharpen the distributions and stabilize the peak near 2.2 M$_\odot$, while the mass-gap assumption shifts it toward higher values around 2.6–2.7 M$_\odot$. In contrast, the radius distributions R$_{\rm TOV}$ show a stronger sensitivity to the underlying hadronic EOS, with typical preferred values around 11.8 km for SFHo+eLSM and 12.4 km for DD2+eLSM when the most robust constraints are applied. In general, the combined constraints consistently favor maximum neutron-star masses near 2.2 to 2.3 M$_\odot$ with radii in the range $12\pm 1$ km. This maximum neutron star masses are consistent with the most massive neutron star currently known, PSR~J0952–0607 with mass $2.35\pm0.17\,M_\odot$ \citep{Romani:2022jhd}. A population analysis yields a conservative lower bound of
$M_{\mathrm{max}}\ge 2.22\,M_\odot$~\citep{Romani:2022jhd}. There is another analysis by Fan et al. \citep{Fan:2024Maxmass}, they obtained very similar value, $2.25_{-0.07}^{+0.08} M_\odot$.

\section*{Acknowledgments}
This work was supported by the Hungarian OTKA fund K138277.
G.K. acknowledges support from the KKP-2026 Research Excellence Programme of MATE, Hungary.

\bibliography{eLSM_bayes}
\bibliographystyle{plain}

\end{document}